# Imprvoing QoS of all-IP Generation of Pre-WiMax Networks Using Delay-Jitter Model

H. Dahmouni[1], H. El Ghazi[1], D. Bonacci[2], B. Sansò[3] and A. Girard[4]

**Abstract**—The topic of this paper is the evaluation of QoS parameters in live Pre-Wimax environments. The main contribution is the validation of an analytical delay-jitter behavior model. These models can be used in optimization algorithms in order to provide opportunistic and reliable all-IP networks. It allows understanding the impact of the jitter constraints on the throughput and packet loss in wireless systems. However, we show that the real-time QoS requirements of real-time and interactive services can be avoided to a large degree by controlling only the packet delay-jitter in a fixed and mobile environment. The QoS metrics have been computed from live measurements in a Pre-Wimax realistic environment (Toulouse/Blagnac Airport).

——————————— ◆ ———————————

## 1 INTRODUCTION

The popularity of interactive and real-time applications has highlighted the limitations of the Internet infrastructure. The different media types exchanged by these applications have significantly different traffic requirements in terms of bandwidth, delay, jitter and reliability, and require different demand service guarantees from the underlying communication network to offer an acceptable performance.

Wireless networks have also seen a tremendous growth in their usage and consequent demand for multimedia applications. The 2nd Generation cellular systems like GSM, which offered circuit-switched voice services, are now evolving towards 3rd Generation (3G) and 4th Generation such as LTE and Pre-WiMax systems that can transmit high-speed data, video and multimedia-traffic [11],[12], [17].

Although ATM has become very popular as the backbone and Backhaul of high-bandwidth networks networks because of the QoS guarantees it can offer, it has not been not been widely accepted as a substitute for the IP stacks used on the Internet. However, the telecommunications industry has been migrating from ATM to IP, ensuring that the IP network can support new services and future traffic growth, and allows users take advantage of low-cost bandwidth from telecommunications providers.

For many future applications, the delay-variation (jitter) remains one of the most important metric of quality of service, and can have a greater impact on the quality than the network latency and packet loss. Even the small amounts of delay variation or jitter introduced by the network can have a significant impact on application performance. With the evolution towards all-IP architecture [11], [13], [21], operators are thinking seriously about dimensioning and optimization tools based on simple and robust delay and jitter models.

Usually, the average network delay and throughput have been used as metrics to optimize the network cost and performance in current design tools [3][4]. This is simply due to the absence of a robust and simple formula for the jitter so that current network planning and design techniques are mostly based on average delay or loss constraints simply because they can often be easily calculated.

As a consequence, the effect of jitter on network structure and operation is not well understood. Getting some qualitative understanding of this QoS requirement will be only possible when we have a fast evaluation method for jitter. We present here a first experimental validation of the jitter model presented in [1]. We can use it to gain insight into the impact of the jitter on the network performance. In particular we focus on the impact of the jitter on the other QoS metrics such as: the throughput and the packet loss.

The objective of the paper is the benchmarking, modeling and analyzing the jitter constraint on Pre-WiMax technology for avionics applications in realistic environments. The live measurements of Pre-WiMax QoS parameters are provided under the SCA project (Communication Systems for Avionics). The innovating character of this project lies in the introduction of IP technology into the avionic domain in order to integrate aircrafts directly into the data-processing network of the airline company.

The system considered the SCA project is based on the use of **the Internet and wireless communications, especially** for Pre-WIMAX, instead of the aeronautical private network for the exchange of AOC/AAC (Aeronautical

————————————————
- *H. Dahmouni with the INPT, Telecommunications Department, Rabat, Morocco.*
- *H. El Ghazi is with the INPT, Telecommunications Department, Rabat, Morocco.*
- *D. Bonacci is with TéSA Laboratory, Toulouse, France.*
- *B. Sansò is with Ecole Polytechnique de Montréal, Electrical Engineering Department, CP 6079 succ Centre-Ville Montreal, Qc, Canada.*
- *A. Girard is with INRS-EMT and GERAD, 800, de la Gauchetire O Suite 6900, Montreal, Qc, Montreal, Qc, Canada.*

Operational Control / Aeronautical Administrative Communication) messages between the avionics domain and the ground system [21],[20].

The remaining of the paper is organized as follows. Section II provides a brief overview of Pre-WIMAX technology. The analytical relationship between traffic load, throughput, packet loss and average jitter is given in section III. In section IV, we describe the measurement platform and some live results of QoS parameters in Pre-WiMax environments. The last section draws some conclusions.

## 2 OVERVIEW OF PRE-WIMAX TECHNOLOGY

Pre-WIMAX technology selected for avionics applications has shown rapid progress and has been widely used to help people doing their daily activities. It has been developed to transmit different types of services data, text and video.

Fixed Broadband Wireless Access (BWA) is a promising technology which can offer high speed voice, video and data service up to the customer [19],[17]. Due to the absence of any standard specification, earlier BWA systems were based on standard 802.11h-Hyperlan 3, which allows 1W of emitted power rather than 0.1W of WiFi. This can improve performance in many situations..

The Wireless MAN standard specifies a Medium Access Control (MAC) layer and a set of PHY layers to provide fixed and mobile Broadband Wireless Access (BWA) over a broad range of frequencies. The Pre-WiMAX has adopted IEEE 802.11h-Hyperlan 3 OFDM PHY layer for the equipment manufacturer due to its robust performance in multi-path transmission [14].

The different advantages offered by Pre-WiMax technology such as: coverage, bandwidth, QoS, etc. make it a good choice for the intermediate technology between Wi-Fi and WiMax. Thus, results provided for pre-WiMax can be extended to the broadband technologies previously cited.

## 3. PERFORMANCE AND MODELING

Delay jitter is an important QoS metric for real-time services, such as VoIP traffic and avionics applications. It can result in consecutive packet experiencing excessive delays and/or consecutive packet loss. Both events lead to a marked deterioration in the subjective quality of interactive and real-time services.

Unfortunately, this parameter is difficult to estimate due to data traffic characteristics and the transient queuing effects in the network. In this section we focus on performance behavior of delay and delay-jitter at packet level by means of an approximate analytical solution. Our aim is to provide an analytical expression for the delay-jitter in function of the traffic load, bandwidth and latency. This is done by taking into account the Poisson nature of the background traffic.

There has been much work during the 1990's on the estimation of cell delay jitter for ATM networks. Most of these results are based on queuing systems and assume that the tagged stream is originally periodic or a general renewal process (e.g.,[5][6][7][8][9]). It is calculated for discrete time processes and FCFS multiplexing operation.

In this paper, we adopt the IETF [22] definition of jitter. It is based on the transit delay of successive packets between the entry and the exit network nodes. Let $T_j$ represents the delay experienced by the jth packet going through a queue. The difference of transit time between two consecutive packets of a tagged flow can be written as

$$J_j = T_{j+1} - T_j \tag{1}$$

which can be positive or negative. The average end-to-end delay jitter is then given by the expected absolute value of this random variable

$$J = E[|T_{j+1} - T_j|] \tag{2}$$

We consider a single node with infinite buffer and a FCFS discipline. There are different streams of packets arriving to this node. Define,

- $\lambda$: the total arrival rate
- C: the link bandwidth
- $\rho$: the total traffic load

In [1] we have shown and validated by simulation that the end-to-end jitter of a tagged flow produced by a single node can be approximated by the following formula:

$$J = \frac{1}{C - \lambda} \left( 1 - e^{-\frac{1-\rho}{\rho}} \left( \frac{1-\rho}{\rho} + e^{-\frac{1-\rho}{\rho}} \right) \right) \tag{3}$$

Note that this formula depends on the traffic parameters only through the bandwidth, the arrival rate, and the traffic load.

On the other hand, the relation that connects the loss probability $B_T$, the average overall throughput $X_T$ of the resource, and the arrival rate is given in [2]

$$B_T = \frac{\lambda - X_T}{\lambda} \tag{4}$$

From equations (3) and (4) we conclude the relationship between the arrival rate, the throughput and the loss rate:

1- When the overall throughput **increas**es then the jitter **decreases**.
2- When the overall throughput **increases** then the loss rate **decreases**.
3- When the overall packet loss **increas**es then the jitter **increases**.

These relationships allow understanding and explaining some phenomena, observed in live measurements, which can highly affect the end-to-end network performance. Typically, we can conclude that the jitter has a behavior opposite that of the throughput, and the same behavior as the loss rate.

## 4. MEASUREMENT AND ANALYSIS

In this section, we present some QoS results carried on live measurements. The Pre-WIMAX technology has been tested to measure various parameters such as: jitter, throughput, and packet loss.

### 4.1 Measurements Platform Description

A sectorial antenna (120° aperture), was placed with the AU (Pre-WiMax Access Unit, equivalent to a WiFi Access Point) on the roof of an Airbus building near the airport tracks. A vehicle was moving on the service roads of the airport (all around the tracks), recording logs every second. An application was developed by M3System to record:
- time stamps,
- position (with integrity level),
- performance (Iperf) logs (bandwidth, jitter, lost packets, ...),
- router logs.

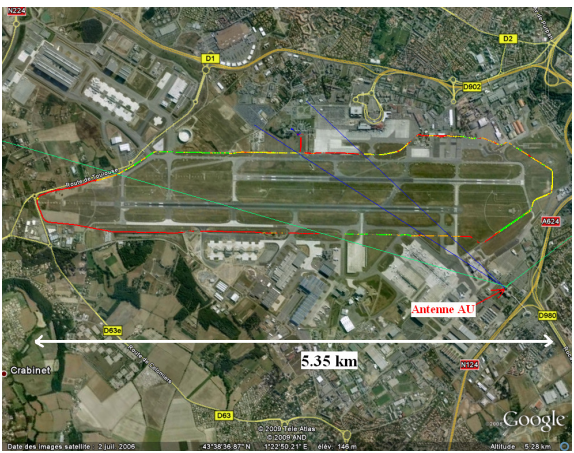

**Figure 1. Airport Pre-Wimax tests conditions**

On Figure 1, the following color code is used:
- Yellow dots → ping OK.
- Green dots → Iperf measurements (TCP or UDP) rate greater than 800kB/s.
- Red dots → ping NOK or rate lesser than 200kB/s.
- Orange dots → intermediate rates.
- No dots → No ping or Iperf logs (means that no signal is received).

The 120° aperture of the antenna is represented by green lines and a masked area (due to another Airbus building) by blue lines. The total width of the Airport is 5.35 km.

### 4.2 Impact of the mobility on communications

As shown in Figure 1, the vehicle was doing round trips at a distance varying from 540 m to 1570 m. TCP and UDP bandwidths were recorded for the test vehicle (shown in Figure 2) moving at different speeds.

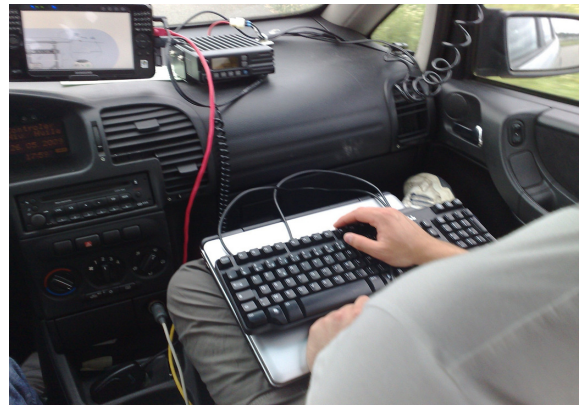

**Figure 2. Measurement Platform**

**a) Measurements for Static Point**

First, we made our live measurements of the quality of service from a static point in order to define the behavior of each parameter of performance and to extract a relationship connecting these QoS parameters.

The measurements for realistic scenarios are investigated to show the impact of distance between Base Station and client station on the QoS parameters of Pre-Wimax technology.

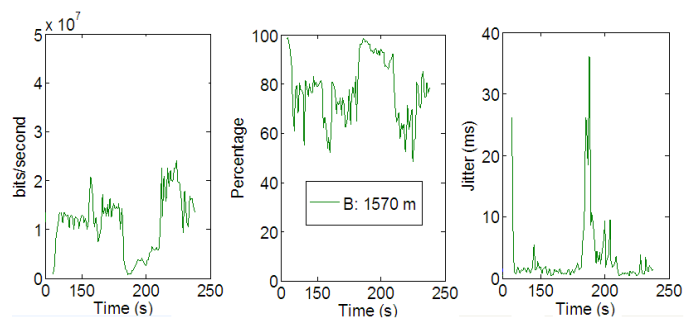

**Figure 3. QoS parameters in static point B**

The QoS metrics are firstly evaluated for a stationary distance of 1570m. At this point, we remark on Figure 3 that end-to-end jitter and throughput are showing an opposite behavior. Periods of high jitter, at around 200 seconds, correspond to low throughput, and vice versa. These fluctuations of QoS parameters are theoretically proved according to the relationship described by analytical model (equation 3 and 4).

**b) Measurements for mobile vehicle with Constant Speed**

Next, the measurements of QoS parameters are performed in mobile environment at a constant speed of 50 km/h. Figure 4 illustrates the measured jitter, throughput and packet loss in this case. As can be seen, the jitter and packet loss variations that are opposite that of the throughput. In fact, regardless of the vehicle speed a high level of jitter can be due to a marked increase in packets loss and to a low throughput level.. We conclude that the vehicle speed has an impact on the QoS parameter values but not on the relationship between these metrics.

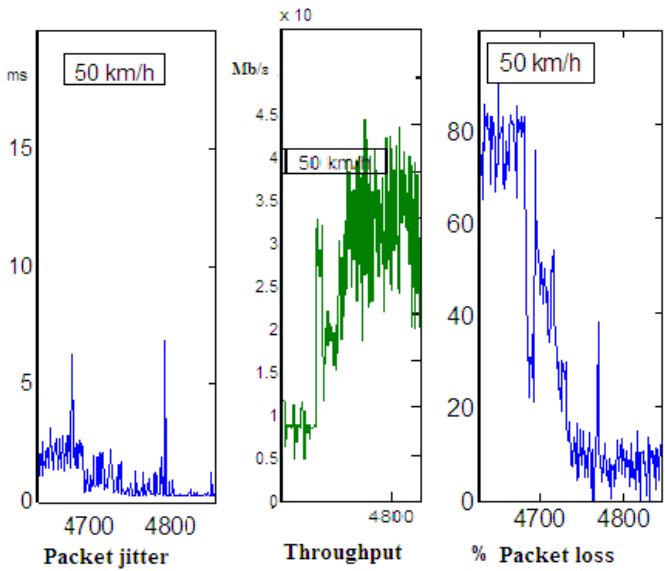

**Figure 4. QoS parameters for 50km/h**

c) **Measurements for variable Speed**

In this section, the measurements results are obtained for different speeds of the vehicle between 10km/h and 50km/h.

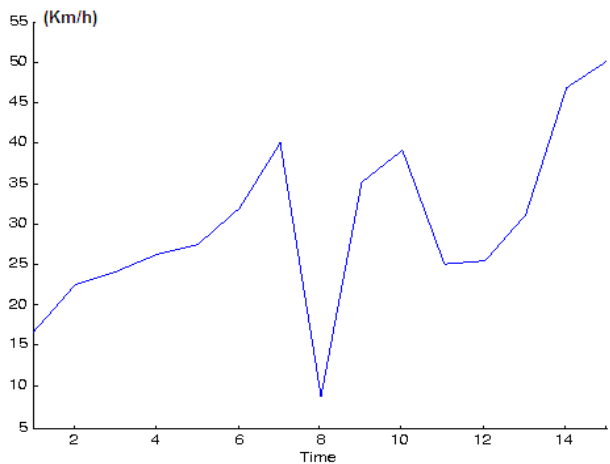

**Figure 5. Speeds versus time**

It is well known that mobility has a significant impact on the performance perceived by mobile stations. This impact is clearly shown within figure 6 and 7 when varying the mobile speed between 10 and 50 km/h. By changing the speed as shown on Figure 5, we can observe a fluctuation of the throughput presented in Figure 6, and the opposite fluctuation for the jitter variation in Figure 7. This is theoretically explained through the relationships connecting jitter, packet loss and throughput.

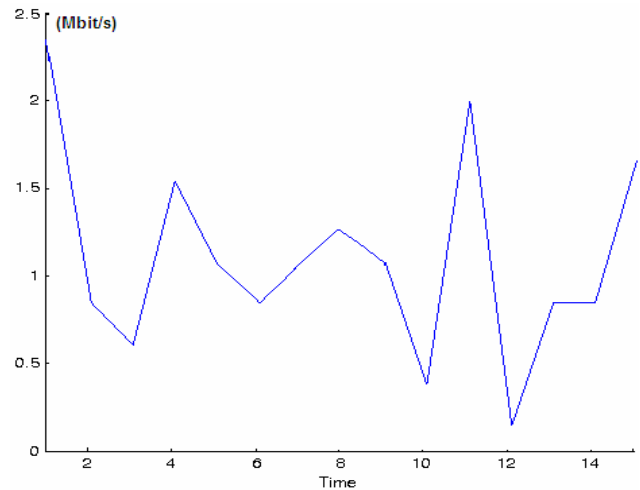

**Figure 6. Throughput variation for different speeds**

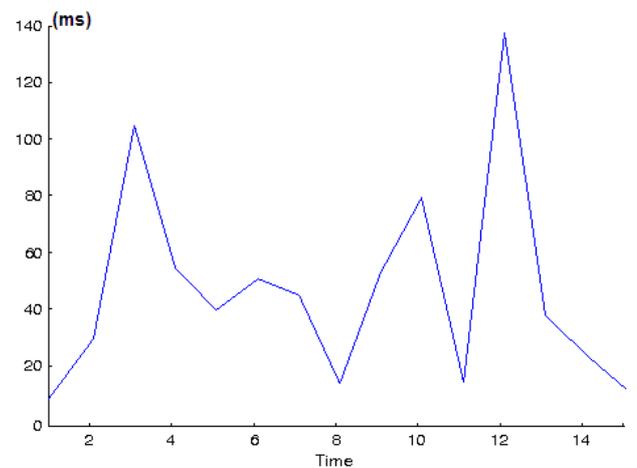

**Figure 7. Jitter variation for different speeds**

## 5. CONCLUSIONS

For Pre-WIMAX technology, the jitter factor impacts the performance of system and Quality of service. In fact, the jitter and packet loss presented the same behavior for static and mobile environments. The qualitative behavior of the jitter and the throughput predicted by the analytical model is confirmed by our live measurements at various speeds. The model can then be used by operator to improve the main factors of quality of service needed by airlines companies. The main application of this relationship can be a component of all-IP network optimization algorithms where the end-to-end jitter, packet loss, throughput and latency appear as a set of QoS constraints.

**ACKNOWLEDGMENTS**

The measurements are provided for SCA project, the authors are indebted to the DGCIS (Direction Générale de la Compétitivité, de l'Industrie et des Services) for its support. The work on the jitter evaluation has been supported by CRD grant CRDPJ 335934- 06 from Canada's National Research Council.

**Hamza Dahmouni** received a PhD degree in Networks and Computer Science from Telecom Bretagne - ENST, France in 2007. He obtained a M. Sc in Networks from Paris-IV University in 2003, and a M. Sc in Wireless network design from Telecom SudParis - INT in 2002. In 2004-2007, he worked as research engineer at France Telecom. His research interests are related to traffic engineering and performance evaluation in heterogeneous networks.

**Hassan El Ghazi** received a PhD degree in electrical engineering from Telecom Lille - UVHC, France in 2008. He obtained a M. Sc in Wireless communications from UVHC in 2004. In 2007-2009, he worked as research engineer at TeSa. His research interests are related to wireless communications, OFDM and traffic engineering.

**David Bonacci** received engineering degree from ENSEEIHT, Toulouse in 1999, and a Ph.D. from Institut National Polytechnique of Toulouse in 2003. He was a research associate at the engineering school ENSEEIHT in 2004 and 2005 and is now permanent research engineer at TeSA laboratory (Telecommunications for Space and Aeronautic). His research activity is centered on radar, spectral analysis, parametric modelings, subband decomposition and wireless telecommunications.

**Brunilde Sansò** is a full professor of Electrical Engineering at Ecole Polytechnique de Montréal and director of the LORLAB. Her interests are in green networking, performance evaluation, reliability and design of wireless and wireline networks. She is a recipient of several awards and honors, Associate Editor of Telecommunication Systems, and editor of two books on planning and performance.

**André Girard** is honorary professor at INRS-EMT and adjunct professor at Ecole Polytechnique of Montreal. His research interests all have to do with the optimization of telecommunication networks and in particular with performance evaluation, routing, dimensioning and reliability. He has made numerous theoretical and algorithmic contributions to the design of telephone, ATM and IP networks.